# Diffuse Gamma Rays from WIMP Decay and Annihilation

Marc Kamionkowski[†]

*School of Natural Sciences, Institute for Advanced Study, Princeton, NJ 08540 U.S.A.*

ABSTRACT

I discuss contributions to the diffuse gamma-ray background from decay and annihilation of weakly interacting massive particles (WIMPs). I first review the calculation of the cosmological abundance of a WIMP and show that it is simply related to the cross section for annihilation of the WIMP into lighter particles. The diffuse extragalactic background radiation (DEBRA) from WIMP decay is then discussed. I show how observational upper limits to the DEBRA can be used to constrain properties of WIMPs that decay to photons, and I present additional new constraints that unitarity of the annihilation cross section imposes on such particles. I then discuss gamma rays from annihilation of WIMPs in the halo, spheroid, and disk of our galaxy, as well as those from WIMP annihilation in the Large Magellanic Cloud.

April 1994

# 1. Introduction

There are currently many complementary experiments probing the sky for energetic astrophysical photons with energies in the x-ray (keV) to VHE gamma-ray ($10^{10}$ GeV) regime. Most of these observations are focussed on photons from burst or point sources; however, many of the experiments also provide a measurement of the diffuse photon background. It is likely that in the near future, our knowledge of the energy spectrum of the diffuse extragalactic background radiation (DEBRA) [1], as well as the diffuse background from our own Galaxy, will be tremendously improved at almost all wavelengths.

The spectrum of the DEBRA is shown in Fig. 1. It is clear that the DEBRA at energies less than a keV is due to a large variety of astrophysical and cosmological sources. Therefore, it is quite likely that the photon background at energies greater than a keV also comes from a variety of different sources. In addition to traditional astrophysical sources, such as AGN and cosmic rays [2], it is also plausible that exotic sources from the early Universe may also contribute to the gamma-ray background.

The Standard Model of elementary-particle interactions is in excellent agreement with all known laboratory experiments. Still, there is almost universal agreement among particle physicists that the Standard Model is incomplete. Suffice it to say that there is no shortage of intriguing new ideas in particle physics, but the results of current and forthcoming accelerator experiments have little to say about most of these ideas. Fortunately, most theories of new physics lead to potentially observable (and sometimes dramatic) astrophysical consequences.

In these lectures, I will discuss contributions to the diffuse gamma-ray background that arise as a result of new physics. The purpose of this endeavor is twofold: First, observations of the gamma-ray background can be used to constrain parameters in new particle theories. The second and perhaps more interesting reason is that gamma rays may provide an avenue for discovery of new physics.

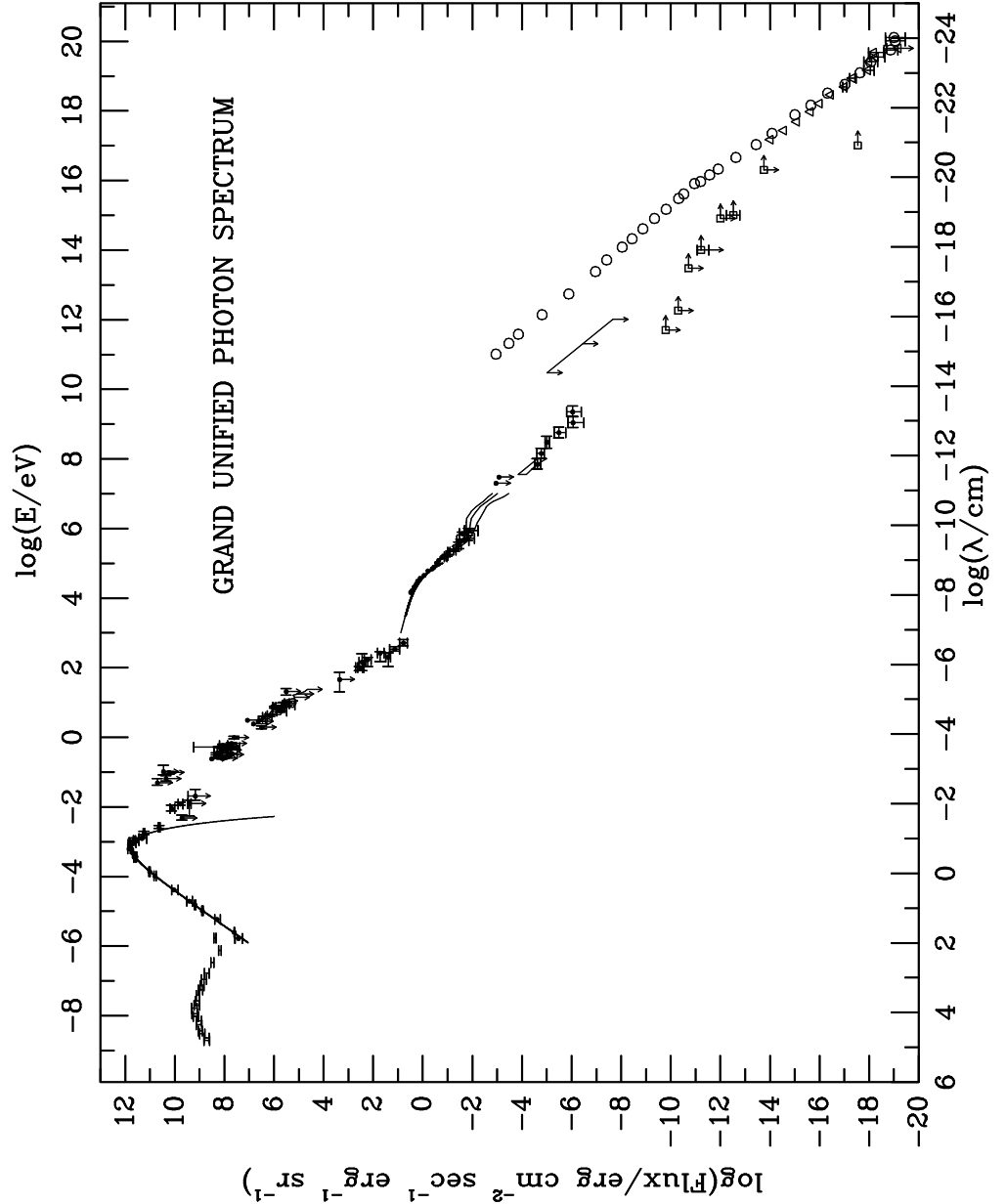

Fig. 1. The diffuse extragalactic photon background. Vertical arrows indicate upper limits and horizontal arrows indicate integrated flux ($> E$). Open circles and triangles indicate the total cosmic-ray flux (gamma rays and hadrons) which places an upper limit to the gamma-ray flux. From [1].

Before continuing, I should mention that there is now evidence that at least part of the diffuse extragalactic x- and gamma-ray background may come from traditional astrophysical sources in the early Universe (for example, quasars and AGN at high redshift) [2]. In addition, there are some recently proposed cosmological models that lead to a diffuse x- and gamma-ray background from $10^6$ solar-mass black holes which collapsed at high redshift [3]. In these talks I will focus on contributions to the diffuse energetic-photon background that may have arisen as a result of processes in the early Universe that involve new physics. There are potentially many new-physics contributions to the diffuse gamma-ray background (for example, from cosmic strings [4] or primordial black holes [5]), but in these lectures, I will focus specifically on gamma rays from decay and annihilation of weakly-interacting massive particles (WIMPs).

The energies of photons produced by new-physics sources spans many orders of magnitude. Throughout these lectures, the term "gamma ray" will be used to denote a photon with energy $\gtrsim$keV, and I will generally not be careful about discriminating between the traditional classification of x-rays, gamma-rays, ultra-high-energy gamma rays, etc. In addition, the focus of this paper will be to describe various ideas—not to implement them as precisely as possible. I will not use the most recent observational data as carefully as possible, and the quantitative precision will generally be order of magnitude. It should be clear from the discussion how the results can be made more precise.

In the next Section, I will show that the interactions of a WIMP specify its cosmological abundance and review some standard cosmological constraints to WIMP masses, lifetimes, and interactions. I will also present some new arguments, based on unitarity of the annihilation cross section, that constrain the masses and lifetimes of WIMPs. In Section 3, I will discuss gamma rays from WIMP decay and show how upper limits to the flux of diffuse extragalactic gamma rays provide constraints to WIMP properties. I then argue that unitarity implies that the energies of gamma rays from WIMP decay in the early Universe should not exceed 200 GeV. In Section 4, I briefly review the evidence for the existence of exotic dark matter in the Universe and in our galactic halo. In Section 5, I discuss gamma rays from WIMP annihilation in the halo, disk, and bulge of our galaxy, and from the Large Magellanic Cloud. I review two distinct signatures of particle dark matter: (i) a characteristic directional dependence of a gamma-ray flux, and (ii) monochromatic gamma rays from direct annihilation of WIMPs into two photons. Section 6 contains a summary and a few concluding remarks.

## 2. Cosmological Abundance a WIMP

Virtually all particle theories predict the existence of new particles, and in many of these theories, the new particles are stable or very long-lived, and weakly interacting. So, let us suppose that in addition to the known particles of the Standard Model there exists a new, yet undiscovered, stable (or long-lived) weakly-interacting massive particle, $X$. In a thermal bath, the number density of $X$ particles is

$$n_X = \frac{g}{(2\pi)^3} \int f(\mathbf{p}) d^3\mathbf{p}, \qquad (2.1)$$

where $g$ is the number of internal degrees of freedom and $f(\mathbf{p})$ is the familiar Fermi-Dirac or Bose-Einstein distribution. At high temperatures ($T \gg m_X$, where $m_X$ is the mass of $X$), $n_X \propto T^3$ (that is, there are roughly as many $X$ particles as photons), while at low temperatures ($T \ll m_X$), $n_X \simeq g(m_X T/2\pi)^{3/2} \exp(-m_X/T)$ (so their density is Boltzmann suppressed). If the expansion of the Universe was so slow that thermal equilibrium was always maintained, the number of WIMPs today would be exponentially suppressed (essentially, there would be no WIMPs); however, due to the expansion of the Universe, at some point the interactions of the WIMP "freeze out" and a relic abundance of $X$ persists.

At high temperatures ($T \gg m_X$), $X$'s are abundant and rapidly converting to lighter particles and *vice versa* ($XX \leftrightarrow l\bar{l}$, where $l\bar{l}$ are quark-antiquark and lepton-antilepton pairs, and if $m_X$ is greater than the mass of the gauge and/or Higgs bosons, $l\bar{l}$ could be gauge and/or Higgs bosons pairs as well). Shortly after $T$ drops below $m_X$ the number density of $X$'s drops exponentially, and the rate for annihilation of $X$'s, $\Gamma = \langle \sigma_A v \rangle n_X$ (where $\langle \sigma_A v \rangle$ is the thermally averaged total cross section for annihilation of $XX$ into lighter particles times relative velocity $v$), drops below the expansion rate $H$. At this point, the $X$'s cease to annihilate, they fall out of equilibrium, and a relic cosmological abundance remains.

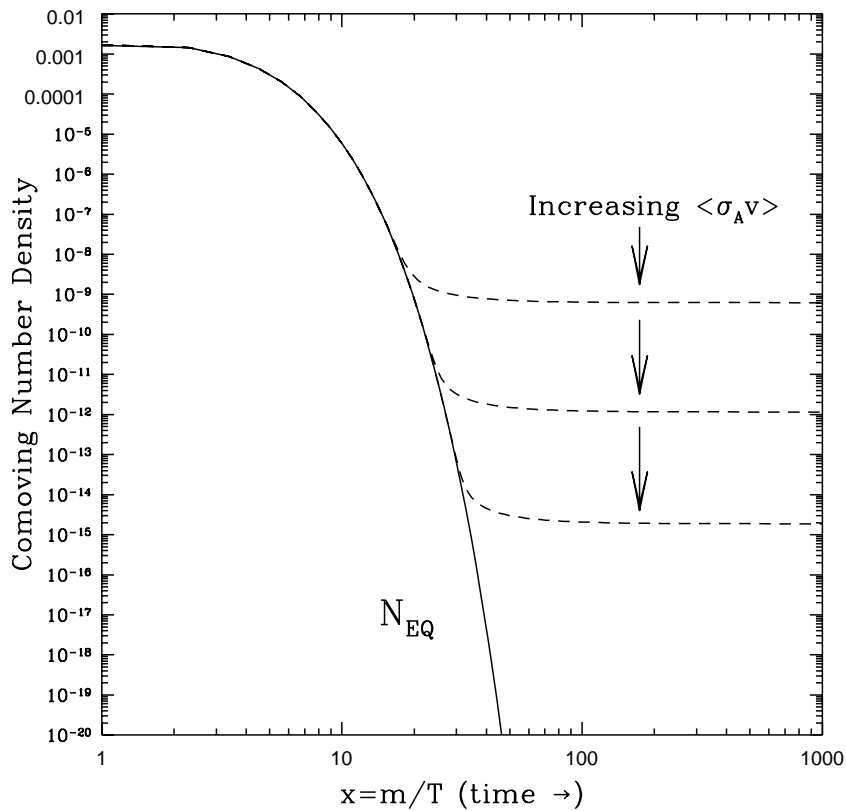

Fig. 2. Comoving number density of a WIMP in the early Universe. The dashed curves are the actual abundance, and the solid curve is the equilibrium abundance. From [6].

In Fig. 2, the equilibrium (solid line) and actual (dashed lines) abundances per comoving volume are plotted as a function of $x \equiv m_X/T$ (which increases with increasing time). As shown in the graph, as the annihilation cross section is *increased* the WIMPs stay in equilibrium longer, and we are left with a *smaller* relic abundance.

Given the annihilation cross section, the relic abundance can be determined quite accurately by solving the Boltzmann equation [6]. Instead of doing so, we will do a simple calculation that actually gives a surprisingly accurate result. The temperature $T_f$ at which the $X$'s freeze out is given by $\Gamma(T_f) = H(T_f)$, where $H(T) = 1.66 g_* T^2/m_{\rm Pl}$ is the expansion rate in the early Universe (that is, the Hubble parameter determined by the Friedmann equation), and $m_{\rm Pl} \simeq 10^{19}$ GeV is the Planck mass. The precise value of $g_*$, the effective number of relativistic degrees of freedom, depends on the temperature, but turns out to be roughly 100 at $T_f$. The freezeout temperature turns out to be $T_f \simeq m_X/25$; there is a small logarithmic dependence on the mass and annihilation cross section. After freezeout, the abundance of $X$'s per comoving volume remains constant. The entropy per comoving volume is also constant, so $n_X/s$ remains constant, where $s \simeq 0.4 g_* T^3$ is the entropy density. So, we use the relations above to find,

$$\left(\frac{n_X}{s}\right)_0 = \left(\frac{n_X}{s}\right)_f \simeq \frac{100}{m_X m_{\rm Pl} g_*^{1/2} \langle \sigma_A v \rangle} \qquad (2.2)$$
$$\simeq \frac{10^{-8}}{(m_X/{\rm GeV})(\langle \sigma_A v \rangle/10^{-27}\,{\rm cm}^3\,{\rm sec}^{-1})},$$

where the subscript $f$ denotes the value at freezeout and the subscript "0" denotes the value today. The current entropy density is $s_0 \simeq 4000$ cm$^{-3}$, and the critical density today is $\rho_c \simeq 10^{-5} h^2$ GeV cm$^{-3}$, where $h$ is the Hubble constant in units of 100 km/sec/Mpc, so the present mass density in units of the critical density, $\Omega_X^0$, is given by,

$$\Omega_X^0 h^2 = \frac{m_X n_X}{\rho_c} \simeq \left(\frac{3\times 10^{-27}\,{\rm cm}^3\,{\rm sec}^{-1}}{\langle \sigma_A v \rangle}\right). \qquad (2.3)$$

The result is independent of the mass of the WIMP (except for logarithmic corrections), and is inversely proportional to its annihilation cross section. It should be noted that $\Omega_X^0$ is the density in WIMPs only if the WIMP is stable (or has a lifetime that exceeds the age of the Universe). If the WIMP decays with a lifetime less than the age of the Universe, then $\Omega_X^0$ is the density the WIMPs *would* have had today if they had *not* decayed.

To summarize, given a particle-physics theory with a stable WIMP, the prescription for determining the abundance of the WIMP is straightforward: calculate the cross section for annihilation of the WIMP into lighter particles, and then the relic abundance is given roughly by Eq. (2.3). If needed, a more accurate result for the abundance can be found by solving the Boltzmann equation.

Before continuing, let us review some general constraints on the masses and lifetimes of WIMPs imposed by the age of the Universe and by partial-wave unitarity of the annihilation cross section [7]. A conservative lower bound on the age of the Universe, $t_0 \gtrsim 10$ Gyr, implies that $\Omega h^2 \lesssim 1$. In addition, it can be shown that unitarity requires that

$$\langle \sigma_A v \rangle \lesssim \frac{3 \times 10^{-22} \text{ cm}^3 \text{ sec}^{-1}}{(m_X/\text{TeV})^2}. \quad (2.4)$$

First, consider the case where the WIMP is absolutely stable or has a lifetime greater than the age of the Universe. Then, Eq. (2.3) and $\Omega_X^0 h^2 \lesssim 1$ tell us that WIMPs with an annihilation cross section $\langle \sigma_A v \rangle \lesssim 3 \times 10^{-27}$ cm$^3$/sec are cosmologically inconsistent. The unitarity bound on the annihilation cross section then implies that $\Omega_X^0 h^2 \gtrsim (m_X/300 \text{ TeV})^2$ for any WIMP (stable or unstable). This tells us that the mass of a stable WIMP must be $m_X \lesssim 300$ TeV in order to be cosmologically consistent [7]. Before moving on, note that if the WIMP decays with a lifetime $\tau \lesssim t_0$ into a massive particle with mass $m_D$, then the mass density of the decay product is $(m_D/m_X)\Omega_X$, and the bounds are diluted accordingly.

Now consider the case where the WIMP decays with a lifetime $\tau \lesssim t_0$ into particles $l$ light enough that they are still relativistic today: ($X \to ll$). The energy density in a nonrelativistic species decays as $R^{-3}$, where $R$ is the scale factor of the Universe, while the energy density in a relativistic species decays as $R^{-4}$ (the additional factor of $R$ is due to the redshifting of the particle energy). Now let us assume that the Universe is flat and matter dominated; if not, then it is straightforward to modify the analysis below, and the qualitative results remain the same. The redshift $z_D$ at which the particle decays is related to the lifetime by $1 + z_D = (\tau/t_0)^{-2/3}$, and in a flat Universe, $t_0 = 2 \times 10^{17}/h$ sec. Therefore, if a WIMP decays at a redshift $z_D$, then the energy density contributed by the decay products is $\Omega_D \simeq \Omega_X^0/(1+z_D)$. Therefore, the annihilation cross section must be $\langle \sigma_A v \rangle \gtrsim 3/(1+z_D)$ cm$^3$/sec, and the unitarity constraint to the mass is $m_X \lesssim 300(t_0/\tau)^{1/3}$ TeV.

Of course it is conceivable that these mass and lifetime limits could be evaded. For example, if there was some non-equilibrium process after freeze-out which produced a significant amount of entropy (such as a first-order phase transition or out-of-equilibrium decay of a massive particle), then the cosmological relic abundance of the WIMP would be diluted, and the unitarity bounds could be avoided.

## 3. Diffuse Extragalactic Gamma Rays from WIMP Decay

Now consider the case where one of the decay products is a photon, $X \to l\gamma$, where $l$ is some noninteracting species [6][1][8]. Actually, it can easily be verified that the conclusions we reach below apply even if $l$ is massive. As above, the fraction of critical density contributed by the decay photons is $\sim (1/2)\Omega_X^0/(1+z_D)$. Let us calculate this result more carefully; in doing so we will also obtain the differential energy spectrum of the decay photons. For now, we will assume that the Universe is transparent to the decay photons. Afterwards, we will discuss where this assumption is valid.

The equation for the time evolution of the energy density $\rho_X$ in the decaying WIMP is

$$\rho_X(R) = \rho_X^0 \left(\frac{R_0}{R}\right)^3 e^{-t/\tau}, \quad (3.1)$$

where $R(t)$ is the scale factor of the Universe as a function of time $t$, $R_0$ is the scale factor today, and again, $\rho_X^0$ is the energy density contributed by the $X$'s today if they had *not* decayed. The factor of $(R_0/R)^3$ in Eq. (3.1) accounts for the decrease in the density due to the expansion of the Universe, and the exponential accounts for the decrease in the density due to decays. The equation for the time evolution of the energy density in decay photons is

$$\dot{\rho}_\gamma + 4H\rho_\gamma = \frac{\rho_x}{2\tau}, \quad (3.2)$$

where $H = \dot{R}/R$ is the expansion rate, and the dot denotes a derivative with respect to $t$. The second term on the left-hand side of Eq. (3.2) accounts for the expansion of the Universe; if we set the right-hand side to zero, we would find $\rho_\gamma \propto R^{-4}$, which is the proper scale-factor dependence of the energy density of a massless species. The right-hand side of Eq. (3.2) accounts for the injection of photons from WIMP decay.

If the Universe is flat and matter dominated, then $R(t) = R_0(t/t_0)^{2/3}$, and the solution to these equations is

$$\rho_\gamma^0 = \frac{\rho_X^0}{2\tau} \int \left(\frac{t}{t_0}\right)^{2/3} e^{-t/\tau}\, dt. \quad (3.3)$$

We will first treat the case where $1 + z_D \gtrsim 1$ (or $\tau \lesssim t_0$). Then the limits of the integral are effectively 0 and $\infty$, and the result is

$$\rho_\gamma^0 = \Gamma\left(\frac{5}{3}\right) \frac{\rho_X^0}{2(1+z_D)} \simeq \frac{\rho_X^0}{2(1+z_D)}, \quad (3.4)$$

since $\Gamma(5/3) \simeq 1.3 \simeq 1$.

This expression was obtained assuming the Universe is transparent to the decay photons. We will now figure out when this is a good assumption. In general, a number of processes can attenuate gamma rays over cosmological distances, and these have been quantified carefully with fairly elaborate calculations [9]. In keeping with the nature of the discussion here, we will only figure out when attenuation of the photons may become important, and to be conservative, we will presume that such photons are simply absorbed (although this may not be entirely true).

The cross section for scattering of photons off electrons drops from the Thomson cross section for photon energies $E \gtrsim$ MeV. The Universe is transparent to photons with energies higher than this threshold, but if the photon energy exceeds $10^6$ GeV or so, production of electron-positron pairs from gamma-ray scattering with cosmic microwave background (CMB) photons becomes possible. The Universe becomes opaque (with an attenuation length of about a few kpc) to higher-energy gamma rays. This threshold, $10^6$ GeV, is easily understood: CMB photons have energies $\sim 10^{-4}$ eV, and the threshold is simply the gamma-ray energy required for there to be enough energy, $2m_e \sim$ MeV, to produce an electron-positron pair in the center of momentum. Therefore, for the purposes of our discussion, the Universe today is transparent to gamma rays with energies MeV $\lesssim E \lesssim 10^6$ GeV.

Of course, earlier in the Universe at redshift $z$, the CMB-photon energies were higher, roughly $10^{-4}(1+z)$ eV, so the threshold for pair production at a given redshift is $10^6(1+z)^{-1}$ GeV. In addition, the energy of a gamma ray whose energy today is $E$ was $E(1+z)$ at a redshift $z$. Therefore, gamma rays with energies $E_0 \gtrsim 10^6(1+z_D)^{-2}$ GeV today would have been attenuated if they were produced at a redshift $z_D$. As Eq. (3.3) makes clear, gamma rays from WIMP decay are not all produced at exactly the same redshift; however, it will become evident in the discussion below that the redshifts of decay of the majority of the WIMPs are closely clustered near $z_D$. Therefore, from now on we will assume that decay photons from WIMPs with masses $m_X/2 \gtrsim 10^6/(1+z_D)$ GeV are attenuated.

The differential energy spectrum of the decay photons is easily obtained from the integral in Eq. (3.3) by noting that the energy of a decay photon today is related to the time at which it decayed (simply from the redshift) by $E(t) = (t/t_0)^{2/3} m_X/2$. The result is

$$\frac{d\rho_\gamma}{dE} = \frac{3}{4} \frac{\rho_X^0}{1+z_D} \left(\frac{E}{E_0}\right)^{5/2} \frac{1}{E} \exp\left[-(E/E_0)^{3/2}\right], \quad (3.5)$$

where

$$E_0 = \frac{1}{2}\frac{m_X}{1+z_D} = \frac{m_X}{2}\left(\frac{\tau}{t_0}\right)^{2/3}. \quad (3.6)$$

The number spectrum of gamma rays ($d\mathcal{F}/d\Omega dE$ is the number flux of gamma rays through a solid angle $d\Omega$ with energies between $E$ and $dE$) is then

$$E\frac{d\mathcal{F}}{d\Omega dE} = \frac{1}{4\pi}\frac{d\rho_\gamma}{dE}$$
$$\simeq 5 \times 10^6 \frac{\Omega_X^0 h^2}{1+z_D}\frac{1}{(E/\text{MeV})}\left(\frac{E}{E_0}\right)^{5/2}\exp\left[-(E/E_0)^{3/2}\right]\text{ cm}^{-2}\text{ sec}^{-1}\text{ sr}^{-1},$$
(3.7)

where $\Omega_X^0$ is again the mass density the WIMPs would have today if they had *not* decayed. Notice that the differential spectrum depends essentially on two parameters: the overall amplitude depends on $\Omega_X^0 h^2/(1+z_D)$, and the energy at which the spectrum is peaked depends on $E_0$. The integrated number flux is then easily obtained:

$$\frac{d\mathcal{F}}{d\Omega} = \frac{1}{8\pi}\frac{\rho_X^0}{1+z_D}\frac{1}{E_0}$$
$$\simeq 10^7 \frac{\Omega_X^0 h^2}{1+z_D}\frac{\text{MeV}}{E_0}\text{ cm}^{-2}\text{ sec}^{-1}\text{ sr}^{-1}.$$
(3.8)

Fig. 3 shows measurements of and upper limits to the diffuse extragalactic background radiation (from Ref. [1]). The solid curves centered at $10^6$ and $10^{12}$ eV are spectra of photons from WIMP decay with $E_0 = 1$ MeV and $10^6$ MeV with $\Omega_X^0 h^2/(1+z_D) = 2 \times 10^{-7}$. If we raise (lower) $\Omega_X^0 h^2/(1+z_D)$, the predicted curve retains the same shape, but is raised (lowered) on this graph. The shape of the curve also remains the same, but the curve is moved to the right (left) as $E_0$ is raised (lowered). If decay photons exist in observable numbers, they would most likely appear as a bump above background at some energy $E$. It is amusing to speculate that the MeV bump in the extragalactic gamma-ray background [10] is actually due to photons from WIMP decay.

The dotted line in Fig. 3 is an approximate upper bound to the flux of diffuse extragalactic gamma rays. We can translate this into an upper bound to the integrated flux of diffuse extragalactic gamma rays with energies near $E$ [6]:

$$\frac{d\mathcal{F}}{d\Omega} \lesssim \frac{\text{MeV}}{E}\text{ cm}^{-2}\text{ sec}^{-1}\text{ sr}^{-1}.$$
(3.9)

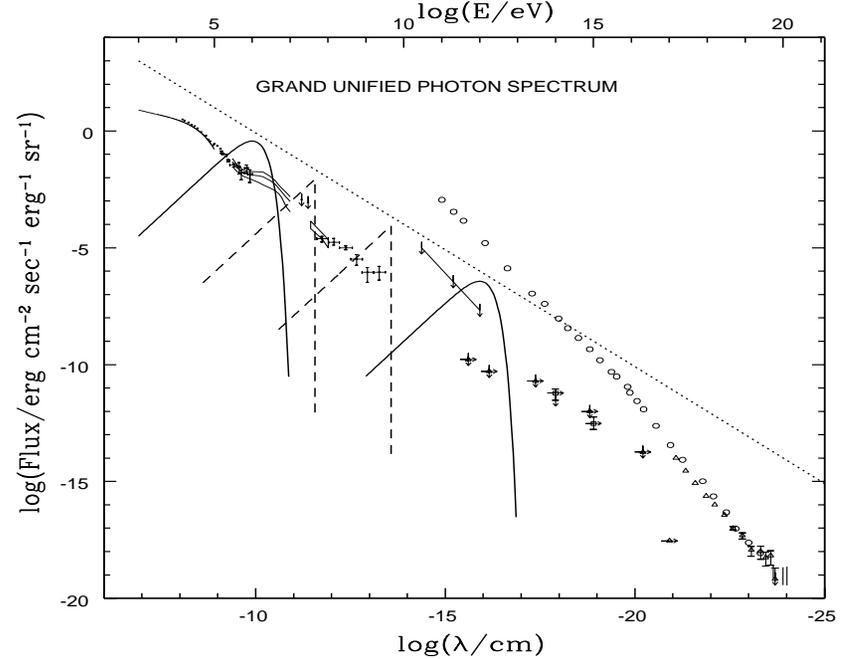

Fig. 3. The diffuse extragalactic gamma-ray background. The dotted curve is an approximate upper bound to the gamma-ray flux. The solid curves are the expected fluxes from a WIMP that decays in the early Universe with $\Omega_X^0 h^2/(1+z_D) = 2 \times 10^{-7}$ and $E_0 = 1$ MeV (at lower energy) and $E_0 = 10^6$ MeV (at higher energy). The dashed curves are the fluxes expected from WIMPs that decay today for $\Omega_X h(\tau/3 \times 10^7 \text{ sec})^{-1}(m_X/\text{GeV})^{-1} = 4 \times 10^{-7}$ ($4 \times 10^{-9}$) and $m_X = 100$ MeV ($10^4$ MeV) for the curve at lower (higher) energies.

Comparing with Eq. (3.8), we find an upper bound to the abundance of a WIMP that decays to a noninteracting particle $l$ and a photon at a redshift $z_D$:

$$\frac{\Omega_X^0 h^2}{1+z_D} = \Omega_X^0 h^2 \left(\frac{\tau}{t_0}\right)^{2/3} \lesssim 10^{-7},$$
(3.10)

and remember that this bound is valid only for particles with masses $m_X/2 \lesssim 10^6(1+z_D)^{-1}$ GeV. Also note that this bound holds only for WIMPs that decay after decoupling but before today, $1 \lesssim z_D \lesssim 1000$. (The inequality involving

11  12

the lifetime $\tau$ is strictly valid only if the Universe is flat and matter dominated. The inequality involving $z_D$ is independent of the geometry of the Universe.) We now recall the relation, Eq. (2.3), between the cosmological abundance and the annihilation cross section, $\langle \sigma_A v \rangle$, and find

$$\frac{\langle \sigma_A v \rangle}{10^{-27}\,\text{cm}^3\,\text{sec}^{-1}} \gtrsim 10^7 (1+z_D)^{-1}. \qquad (3.11)$$

It is interesting to note that if the WIMP decays to a photon with a lifetime in the range considered here, the bound on the abundance due to gamma-ray constraints is *much* stronger than the age-of-the-Universe bound, $\Omega_X^0 h^2 \lesssim (1+z_D)$. In addition, particles which decay to photons with cosmological lifetimes have extraordinarily weak couplings to photons. There are no traditional laboratory experiments which probe such weak interactions, so astrophysical information—like gamma-ray observations—provide invaluable constraints to hypothetical particles.

Now let us turn once more to the constraints to the cosmological abundance imposed by unitarity and investigate the implications for extragalactic gamma rays from WIMP decay. Eqs. (2.4) and (3.11) imply that if the WIMP decays to a photon after decoupling of CMB photons but before today, then then the WIMP mass must be $m_X \lesssim 200(1+z_D)^{-1/2}$ GeV (unless $m_X/2 \gtrsim 10^6/(1+z_D)$ GeV in which case the photons would have been attenuated and this bound would be evaded). Since $E_0 = m_X/2(1+z_D) \lesssim 100(1+z_D)^{-1/2}$ GeV, the maximum energy of gamma rays from WIMPs that decay with lifetimes $10^{12}\,\text{sec} \lesssim \tau \lesssim 10^{17}\,\text{sec}$ is about 100 GeV. It is encouraging to note that *if* a gamma-ray signature of such WIMPs does exist, it should most likely fall in the energy range accessible to EGRET [11]. We should caution that this bound applies only to the simplest (and most plausible) models. There may of course be loopholes in these arguments (such as production of entropy after freezeout of the WIMP), and this limit applies only to relics that are produced thermally in the early Universe. Gamma rays from decaying particles produced by some other mechanism could conceivably have higher energies.

Before concluding this Section, let us complete the discussion by working out the spectrum of gamma rays from WIMPs that decay at the present epoch—that is, WIMPs with lifetimes comparable to or greater than the age of the Universe, $\tau \gtrsim 2 \times 10^{17} h^{-1}$ sec. In this case, most of the majority of the WIMPs have not yet decayed, and the energy density in decay photons is small compared to the current mass density in WIMPs. We return to Eq. (3.3), but now the upper limit on the integral is $t_0$. Again recalling that the energy of a gamma ray produced by a WIMP decay at time $t$ is $E(t) = m_X(t/t_0)^{2/3}/2$, and that the age of the Universe is $t_0 = (2/3)H_0^{-1}$, it is easily shown that

$$\frac{d\rho_\gamma}{dE} = \frac{\rho_X(R_0)}{\tau H_0 m_X}\left(\frac{E}{m_X/2}\right)^{3/2} \quad \text{for } E \leq \frac{m_X}{2}. \qquad (3.12)$$

Plugging in the numbers, the differential gamma-ray flux turns out to be

$$E\frac{d\mathcal{F}}{d\Omega dE} = \frac{\rho_X(R_0)}{4\pi\tau H_0 m_X}\left(\frac{E}{m_X/2}\right)^{3/2}$$
$$\simeq 2.4 \times 10^4 \frac{\Omega_X h}{(\tau/3 \times 10^{17}\,\text{sec})(m_X/\text{GeV})}\left(\frac{E}{m_X/2}\right)^{3/2}\,\text{cm}^{-2}\,\text{sec}^{-1}\,\text{sr}^{-1}. \qquad (3.13)$$

The dashed curves in Fig. 3 are the fluxes expected from WIMPs that decay today for $\Omega_X h(\tau/3 \times 10^7\,\text{sec})^{-1}(m_X/\text{GeV})^{-1} = 4 \times 10^{-7}$ ($4 \times 10^{-9}$) and $m_X = 100$ MeV ($10^4$ MeV) for the curve at lower (higher) energies. Again, the gamma-ray spectrum from WIMP decay at the present epoch can generally be described by two parameters: (i) an amplitude, which is proportional to the WIMP density and inversely proportional to the WIMP lifetime and mass, and (ii) the gamma-ray energy, which depends only on the WIMP mass.

The integrated gamma-ray flux turns out to be

$$\frac{d\mathcal{F}}{d\Omega} = 1.6 \times 10^4 \frac{\Omega_X h}{(\tau/3 \times 10^{17}\,\text{sec})(m_X/\text{GeV})}\,\text{cm}^{-2}\,\text{sec}^{-1}\,\text{sr}^{-1}. \qquad (3.14)$$

The upper limit to the DEBRA, Eq. (3.9), can then be used to place the following upper bound to the cosmological abundance of a WIMP that decays to a photon with a lifetime greater than the age of the Universe:

$$\Omega_X h \lesssim 10^{-7}(\tau/3 \times 10^{17}\,\text{sec}) \qquad \text{for} \qquad \tau \gtrsim 3 \times 10^{17}\,\text{sec}. \qquad (3.15)$$

The unitarity constraint on the cosmological abundance, $\Omega_X h^2 \gtrsim 10^{-5}(m_X/\text{TeV})^2$, then tells us that $m_X \lesssim 100(\tau H_0)^{1/2}$ GeV for a WIMP that decays to photons with a lifetime greater than the age of the Universe. Again, these constraints are *much* stronger than the age-of-the-Universe constraints if the WIMP lifetime falls in the range $t_0 \lesssim \tau \lesssim 10^6 t_0$. As a check, note that our results for WIMPs with $\tau \gtrsim t_0$ agree with those we obtained previously for $\tau \lesssim t_0$ when $\tau \simeq t_0$.

If the WIMP lifetime is greater than the age of the Universe, then it is likely that WIMPs cluster with galaxies, and the angular distribution of gamma rays from WIMP decay may be anisotropic, with a stronger signal from galaxies or clusters of galaxies. In particular, long-lived WIMPs should cluster in our own galactic halo, and the gamma-ray signal from WIMP decay in the halo could conceivably dominate that from extragalactic sources. In this case, the diffuse gamma-ray background from WIMP decay will have an angular distribution that depends on the distribution of dark matter in our halo. Similar arguments have been developed for the angular distribution of gamma rays from WIMP annihilation in the halo, as discussed in Section 5.

Before we conclude this discussion of diffuse extragalactic gamma rays from WIMP decay into photons, we point out that in many models the WIMP may decay into some other final state, such as quarks or leptons, with a branching ratio significantly greater than that for decay into photons. If the WIMP decays primarily into some other final state, the simple analysis above does not apply directly; however, there may still be reason to expect a gamma-ray signal from such decays. For example, if a WIMP decays primarily to quarks, the quarks will shower and produce hadronic jets similar to those observed in accelerator experiments. Neutral pions in the shower will decay to photons. The energy spectrum of such photons will be rather broad, and centered at energies roughly an order of magnitude smaller than the WIMP mass [12]. For a given decay channel, calculation of the cosmic gamma-ray flux is then straightforward, although slightly more complicated than in the case where the WIMP decays directly into photons.

Throughout this Section we assumed the Universe was flat and matter dominated. It is straightforward to re-do the calculations if the Universe is open or closed, or perhaps flat but dominated by a cosmological constant, and our rough limits on lifetimes and masses will be altered perhaps by factors no greater than a few. Although cosmic gamma rays from WIMP decay are not a generic result of most theories with new particles, it is interesting to note that observational information on the diffuse extragalactic gamma-ray background can be used to constrain hypothetical long-lived particles which decay either directly or indirectly into photons.

## 4. Dark Matter and WIMPs

In the previous Section, we considered gamma rays from decaying WIMPs. In fact, what is far more interesting (and perhaps likely) is the case where the WIMP is absolutely stable, or at least has a lifetime much greater than the age of the Universe. If so, WIMPs can provide a solution to the dark-matter problem. Although we would not necessarily expect gamma rays from WIMP decay in this case, there could be a distinctive diffuse background of gamma rays from WIMP annihilation, as we will discuss below.

One of the most intriguing problems in cosmology is the nature of the dark matter [13]. There are many reasons to believe that an overwhelming fraction of the mass density in the Universe is nonluminous ($\Omega \gg \Omega_{LUM} \simeq 0.01$). If the Universe currently has $\Omega$ of order unity (to within a few orders of magnitude), then the value of $\Omega$ at the Planck epoch must have been $1\pm 10^{-60}$. Aesthetically, this suggests that the Universe is flat, in which case $\Omega = 1$ today. Another reason to believe the Universe is flat is that the most promising solution to the horizon problem, inflation, sets $\Omega$ to unity. The existence of structure in the Universe provides another theoretical argument that there is far more matter than is seen: Cosmological density perturbations grow only in a matter-dominated Universe; if the luminous matter was all there was, the era of matter domination would have been very short thereby requiring initial perturbations which would have resulted in anisotropies in the cosmic microwave background radiation much larger than those observed [14].

Theoretical arguments aside, there is ample observational evidence for the existence of dark matter. By applying Newton's laws to the motion of galaxies in clusters, one infers a mass density in clusters of $\Omega \simeq 0.1$–$0.3$. Results from gravitational lensing seem to confirm this. Observers attempting to measure the mass density of the Universe from observed peculiar velocities find values of $\Omega$ near unity and certainly no less than 0.3 [15].

The most convincing evidence involves galactic dynamics. There is simply not enough luminous matter to account for the observed rotation curves of galaxies. From its gravitational effects, one infers a galactic dark halo of mass 3-10 times that of the luminous component. In particular, dark matter must exist in our own galactic halo, and the density of dark matter in the local neighborhood can be determined to be about 0.3 GeV cm$^{-3}$ to about a factor of two [16].

The resulting question has a certain intrinsic grandeur and simplicity: What is the Universe made of? What about baryons? In order to reproduce the observed abundances of light nuclei, big-bang nucleosynthesis (BBN) requires a baryon density of $\Omega_b h^2 \simeq 0.01$ [17]. So, there is reason to believe that some baryons are dark, perhaps in the form of massive compact halo objects (MACHOs), though probably not enough to fill galactic halos. However, there are no good reasons for these baryons to remain dark, nor are there any plausible mechanisms for them to wind up in the halo. The recent observation of microlensing events [18] suggests that there are indeed MACHOs; however, it is not yet clear if the rate of microlensing events is sufficient to account for the halo dark matter. Preliminary results seem to indicate that the rate of lensing events in the direction of the galactic center is much higher than that in the direction of the Large Magellanic Cloud, suggesting that the observed MACHO events may all be in the disk, and not necessarily in the halo. At this point, the statistics are still preliminary; a more definitive result should be available in the forthcoming years. Even if there were some baryonic dark matter, there is no way, according to BBN, that it could account for the dynamics of clusters, let alone contribute values of $\Omega$ closer to unity.

Another dark-matter candidate in the Standard Model is a neutrino with mass $\mathcal{O}(30\,\text{eV})$. Although a possibility, N-body simulations of structure formation in a neutrino-dominated Universe do a poor job of reproducing the observed structure in the Universe [19]. Furthermore, it is difficult to see (essentially from the Pauli principle), how such a neutrino could make up the dark matter in galaxies.

Thus we are lead to consider the possibility that the dark matter consists of some undiscovered elementary particle. There are many ideas for new physics beyond the Standard Model which predict the existence of new stable particles which have weak interactions with matter. In particular, the most well developed idea along these lines is supersymmetry, which predicts the existence of a plethora of new particles, the lightest of which is stable and weakly interacting. In most cases, this stable particle is the neutralino, a linear combination of the photon, $Z$ boson, and Higgs bosons. Depending on its specific composition, the neutralino is sometimes referred to as a photino, $B$-ino, gaugino, or higgsino. Other WIMP candidates which have been considered in the literature are massive Dirac or Majorana neutrinos, although Dirac neutrinos have now been ruled out by a variety of experiments as the dark matter in the halo [20].

In Section 2, we showed that if such a particle exists and is stable, its relic abundance is $\Omega_X h^2 \simeq 3 \times 10^{-27}\,\text{cm}^3\,\text{sec}^{-1}/\langle \sigma_A v \rangle$. A simple dimensional estimate for the order of magnitude expected for an annihilation cross section for a particle with weak-scale interactions is $\langle \sigma_A v \rangle \sim \alpha^2/(100\,\text{GeV})^2 \sim 10^{-25}\,\text{cm}^3\,\text{sec}^{-1}$, for $\alpha \sim 10^{-2}$. This is remarkably close to the value required to account for the dark matter, especially if we realize that there is no *a priori* reason for a weak-scale interaction to have anything to do with closure density, a cosmological parameter! This simple coincidence—which suggests that if a WIMP exists, it *is* the dark matter—has been followed by extensive theoretical work, and has led to an enormous experimental effort to detect these WIMPS.

Before discussing the possible gamma-ray signatures of WIMPs, we briefly review some of the other possible signatures of WIMP dark matter. Although quite weak, the WIMP has a finite cross section for elastically scattering off of a nucleus. Therefore, one method of detecting WIMPs is by observing the $\mathcal{O}(\text{keV})$ recoil energy imparted to a nucleus in a low-background detector when a WIMP in the halo scatters off of it [21]. Another technique involves searching for energetic neutrinos from the Sun or the core of the Earth in astrophysical neutrino detectors (such as DUMAND, AMANDA, Kamiokande, MACRO, IMB, etc.) [22][23]. WIMPs in the halo will accumulate in the Sun and in the Earth and annihilate therein in much the same way they annihilated in the early Universe. Among the annihilation products will be energetic neutrinos which, if detected, could hardly be confused with anything else. Along similar lines are searches for cosmic rays from WIMP annihilation in the halo [24]. Annihilation in the halo could plausibly produce a distinctive spectrum of low-energy cosmic-ray antiprotons and/or high-energy positrons which do not arise in standard models of cosmic-ray propagation.

## 5. Gamma Rays from WIMP Annihilation

If WIMPs populate our galactic halo, then they may annihilate—in much the same way they did in the early Universe—and produce a diffuse background of gamma rays [12][25][26][27][28][29]. Unlike gamma rays from WIMP decay in the early Universe, these gamma rays will come primarily from WIMPs in our own Galaxy (or possibly the Large Magellanic Cloud) and are not extragalactic in origin. The diffuse background of gamma rays from standard astrophysical sources is poorly understood and precise measurements are difficult; therefore, any inference of the existence of dark matter in the halo from gamma-ray observations must come from fairly distinct gamma-ray signatures. In this Section, I will discuss two such signatures: (i) a distinct feature in the gamma-ray energy spectrum, and (ii) a distinct angular spectrum from WIMP annihilation in the halo. First we will discuss qualitatively the signatures, and then we will make order-of-magnitude estimates of the expected gamma-ray fluxes from simple plausible models for the WIMP.

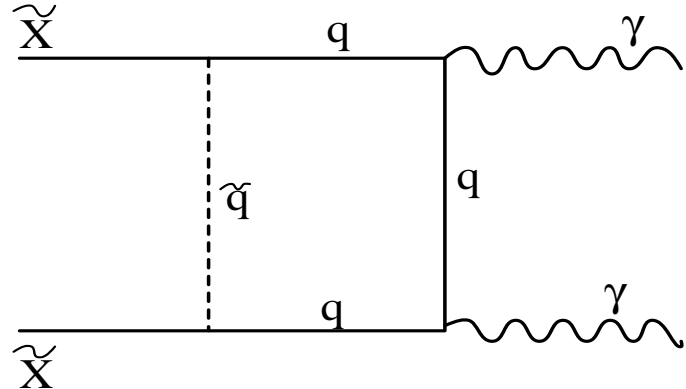

Fig. 4. Example of a Feynman diagram for annihilation of two WIMPs (taken to be neutralinos $\tilde{\chi}$) into photons. The fermions (solid lines) in the loops are quarks (although they could also be leptons), and the scalar (the dotted line) is a squark (slepton), the supersymmetric partner of the quark (lepton).

When WIMPs annihilate to quarks and leptons (and/or Higgs and gauge bosons if the WIMPs are heavy enough) in the galactic halo, the subsequent shower from hadronization of the quarks will produce gamma rays with a broad energy distribution centered roughly around 1/10th the WIMP mass [12]. Such a broad signal will in general be difficult to distinguish from background.

WIMPs, essentially by definition, have no direct coupling to photons. However, by virtue of the fact that the WIMP must have some appreciable coupling to ordinary matter (or else annihilation in the early Universe would be too weak to provide $\Omega_X h^2 \lesssim 1$), it is almost guaranteed that any realistic WIMP will couple to photons through loop diagrams, for example, that shown in Fig. 4. Therefore, there will always be some small, but finite, cross section for direct annihilation of two WIMPs into gamma rays. Since the typical velocity of WIMPs in the halo ($\sim 300$ km sec$^{-1}$) is very small compared with the velocity of light, photons produced by annihilation of WIMPs will be monochromatic at an energy equal to the WIMP mass [28]. No easily imaginable traditional astrophysical source produces monochromatic gamma rays at energies in the range 10-1000 GeV, or so. Therefore, observation of monochromatic gamma rays would provide a "smoking-gun" signal for the existence of WIMPs in the halo.

The second signature of dark-matter annihilation in the halo involves observation of a characteristic angular dependence of the gamma-ray intensity [26]. The simplest model for the density distribution of a dark galactic halo needed to account for the flat rotation curves is an isothermal sphere which has a mass-density distribution,

$$\rho(r) = \rho_0 \frac{R^2 + a^2}{r^2 + a^2}, \tag{5.1}$$

where $r$ is the distance from the center of the galaxy, and $a$ is the scale length of the halo. In our own galaxy, $R \simeq 8$ kpc is our distance to the galactic center, and $\rho_0 \simeq 0.3$ GeV cm$^{-3}$ is the local mass density of dark matter, which is uncertain to within about a factor of two. Also, in our galaxy, $R/a \sim 1$ and is also uncertain to within a factor of two. The asymptotic value of the rotational velocity at large radii fixes the quantity $\rho_0 a^2$, and the uncertainties in $\rho_0$ and $a$ (which are correlated) come from uncertainties that arise in modeling the Galaxy [30]. In addition to accounting for flat galactic rotation curves, isothermal spheres seem to result in N-body simulations of gravitational collapse of noninteracting particles in an expanding Universe. Although it is quite plausible (if not likely) that the halo density profile may differ slightly from an exact isothermal sphere, Eq. (5.1) provide a simple model to work with.

Given an isothermal distribution of dark matter, it is straightforward to calculate the angular dependence of the gamma-ray flux from WIMP annihilation as a function of $\psi$, the angle between the line of sight and the galactic center:

$$\begin{aligned}\frac{d\mathcal{F}}{d\Omega} &= \frac{\sigma_{XX\to\gamma\gamma} v}{4\pi m_X^2} \int_0^\infty \rho^2(r)\, dr(\psi) \\ &\simeq 2\times 10^{-12}\, \frac{(\sigma_{XX\to\gamma\gamma}/\,10^{-30}\,\text{cm}^3\,\text{sec}^{-1})}{(m_X/\,10\,\text{GeV})^2}\, I(\psi),\end{aligned} \tag{5.2}$$

where $r(\psi)$ is the distance along a line of sight at an angular separation $\psi$ from the galactic center. The quantity $\sigma_{XX\to\gamma\gamma}$ is the cross section for annihilation of WIMPs into two photons. Fig. 5 shows the result for $I(\psi)$, the angular dependence of the gamma-ray flux, for three values of the ratio $R/a$. Observation of such a signal would provide intriguing evidence for WIMPs in the halo. To continue the arguments of Section 3, if WIMPs populate the halo and decay to photons, then there will be also be an angular dependence to the gamma-ray intensity. However, the gamma-ray intensity in this case would be proportional to a line integral of the density rather than the square of the density as in Eq. (5.2), so the variation in intensity with angle would not be as dramatic as that from annihilation.

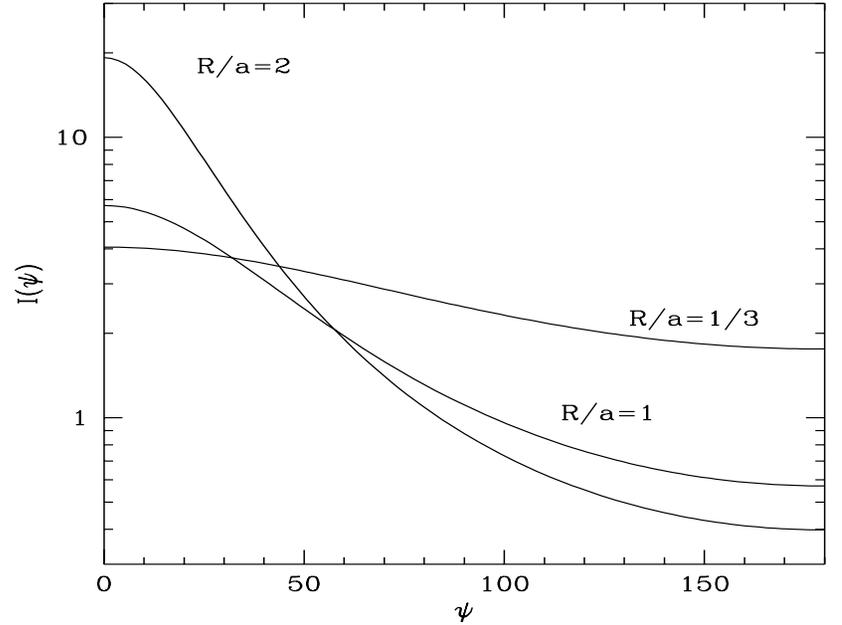

Fig. 5. The intensity of a gamma-ray signal from WIMP annihilation in the halo as a function of the angle $\psi$ between the line of sight and the galactic center. (As in Turner [26].)

Along similar lines, it has been suggested that there may also be an enhancement in the dark-matter density in the galactic bulge or in the disk, and if this dark matter were made of WIMPs, annihilation could lead to a strong gamma-ray signal from the galactic center or the disk [25]; however, it is difficult to see why WIMPs would accumulate in the galactic bulge or in the disk. Recently, Gondolo has suggested that the Large Magellanic Cloud (LMC) could be immersed in a halo of dark matter with a central density 10 times that of our own galaxy, and that annihilation of dark matter therein could lead to a gamma-ray intensity from the LMC roughly ten times stronger than that from our own halo [31], although this estimate comes with significant uncertainties.

Until now, we have discussed only qualitatively some possible signature. We will now do an order-of-magnitude estimate of the expected fluxes. For example, a simple estimate of the cross section that arises from the Feynman diagram for annihilation of two neutralinos $\tilde{\chi}$ into photons shown in Fig. 4 is

$$\sigma_{XX \to \gamma\gamma} v \simeq \frac{\alpha^4 m_X^2}{m_{\tilde{q}}^4}, \tag{5.3}$$

where $m_{\tilde{q}}$ is the mass of the squark $\tilde{q}$ and $\alpha \simeq 10^{-2}$. The squark is generally the heaviest particle in the loop, so the squark propagator leads to a suppression $m_{\tilde{q}}^{-4}$ in the cross section. The factor of $\alpha^4$ in Eq. (5.3) comes from the four couplings in the diagram (which are then squared to give a cross section), and the factor of $m_X^2$ in the numerator must be included to make the cross section dimensionally correct.

Now, for purposes of illustration, let us focus on the case that the WIMP is a pure $B$-ino, a linear combination of the supersymmetric partners of the photon and $Z$ boson which turns out to be the lightest supersymmetric particle in many theories. In this case, the relic abundance turns out to be [32],

$$\Omega_{\tilde{B}} h^2 \simeq 7 \times 10^{-3} \left(\frac{m_{\tilde{q}}}{m_X}\right) \left(\frac{m_{\tilde{q}}}{100 \, \text{GeV}}\right)^2. \tag{5.4}$$

Assuming the Universe is flat and that $B$-inos are the dark matter, then $\Omega_{\tilde{B}} h^2 \simeq 0.25$, and $\sigma_{XX \to \gamma\gamma} \simeq 3 \times 10^{-31}$ cm$^3$ sec$^{-1}$. If we insert this estimate into Eq. (5.2) and compare with measurements at 10 GeV shown in Fig. 1, we conclude that the problem with gamma-ray signatures from dark-matter annihilation is that the signals in generic models, even with optimistic astrophysical assumptions, are at best only marginally observable with current detectors.

On the other hand, it should be remembered that there are significant uncertainties in both the particle physics and astrophysics. For example, the properties of WIMPs vary significantly from model to model, and the cross sections for producing gamma rays in some models may be significantly larger than the simple estimate in Eq. (5.3). There are also indications that heavier WIMPs which couple to the $W^\pm$ boson, such as higgsinos, will annihilate more efficiently into gamma rays [29]. If there is a bulge population of WIMPs, or if WIMPs in the halo are clustered into clumps [33], then the gamma-ray flux from WIMP annihilation could be increased. Finally, there should be substantial improvements in observational high-energy gamma-ray astronomy in the forthcoming years.

## 6. Summary and Conclusions

Gamma-ray astronomy is a rapidly developing field. The recent deployment of orbiting gamma-ray detectors is being complemented by the development of ground-based observatories that will probe higher energies. Although the primary aim of most of these efforts is to detect gamma rays from point sources, information on the diffuse background of gamma rays should also be available.

There are numerous reason to believe that there is new physics beyond the Standard Model of elementary-particle interactions. Many extensions of the Standard Model predict the existence of new weakly-interacting massive particles. We have argued that if such a particle exists, it would have an interesting cosmological abundance. If these particles are unstable and decay either directly or indirectly into photons with lifetimes greater than the age of the Universe at decoupling, they can contribute to the diffuse extragalactic gamma-ray background. Observational limits to the gamma-ray background can be used to provide strong constraints on the masses and lifetimes of such hypothetical particles.

Once again, the discussion of gamma rays from WIMP decay relied on several simplifying assumptions. First, we assumed that gamma rays are unattenuated as they propagate through the intergalactic medium. This is true for gamma rays with an energy $\lesssim 10^6$ GeV today, the energy at which electron-positron pair production by scattering with CMB photons becomes possible. The threshold for gamma rays that originate early in the Universe at high redshift is much lower; in the most extreme case, gamma rays produced at redshifts near decoupling with an energy *today* greater than a GeV could have been attenuated. The other simplifying assumption made is that the Universe is flat and matter dominated. If the Universe is either open or closed, or perhaps flat but vacuum (cosmological-constant) dominated, then the time dependence of the scale factor is different. If so, then our quantitative results may be altered by a factor of a few or so, but certainly much less than an order of magnitude. It should also be remembered that the cosmological abundance of a particle is related to its annihilation cross section only if it is a thermal relic. There are indeed some candidate dark-matter particles (such as the axion) that are produced non-thermally in the early Universe. In these cases, the abundance is not related to the annihilation cross section, and the unitarity limits do not apply.

The most interesting case is that where the WIMP is stable, or at least long-lived compared with the age of the Universe. If so, the WIMP can naturally account for the dark matter known to exist throughout the Universe and in our own galactic halo. In this case, WIMP annihilation in the halo can plausibly lead to a distinct and observable background of diffuse gamma rays. Observation of monochromatic gamma rays with an energy between a GeV and a TeV would provide very convincing evidence for the existence of dark matter in our halo. Another possible signature would be observation of an angular variation of the gamma-ray intensity that traces the distribution of dark matter in our halo, or possibly in the galactic center or LMC.

Unfortunately, ballpark estimates of the gamma-ray flux expected from generic models of WIMP annihilation fall far below current observational sensitivities. On the other hand, the specific properties of the WIMP vary considerably from one model to the next. Also, experimental advances, driven primarily by curiosity about traditional astrophysical sources, will continue. Therefore, there is always a chance that a gamma-ray signal indicating the existence of WIMPs is "just around the corner." Given the tremendous impact such a discovery would have on physics and cosmology, perhaps this is worth keeping in mind.

I am grateful to Ted Ressell for providing the DEBRA figure. I would like to thank the organizers for inviting me to lecture at this very enjoyable School. This work was supported by the U.S. Department of Energy under contract DEFG02-90-ER 40542.